# Volatile and Nonvolatile Resistive Switching in Lateral 2D Molybdenum Disulfide-Based Memristive Devices

*Sofía Cruces[#], Mohit D. Ganeriwala[†], Jimin Lee[#], Ke Ran[¥,§,Ω], Janghyun Jo[Ω], Lukas Völkel[#], Dennis Braun[#], Bárbara Canto[¥], Enrique G. Marín[†], Holger Kalisch[‡], Michael Heuken[‡,¶], Andrei Vescan[‡], Rafal Dunin-Borkowski[Ω], Joachim Mayer[§,Ω], Andrés Godoy[†], Alwin Daus[∥], and Max C. Lemme[#,¥\**

[#] Chair of Electronic Devices, RWTH Aachen University, Otto-Blumenthal-Str. 25, 52074 Aachen, Germany.

[†] Department of Electronics and Computer Technology, Facultad de Ciencias, Universidad de Granada, 18071, Granada, Spain.

[¥] AMO GmbH, Advanced Microelectronic Center Aachen, Otto-Blumenthal-Str. 25, 52074 Aachen, Germany.

[§] Central Facility for Electron Microscopy, RWTH Aachen University, Ahornstr. 55, 52074, Aachen, Germany.

[Ω] Ernst Ruska-Centre for Microscopy and Spectroscopy with Electrons (ER-C), Forschungszentrum Jülich GmbH, Wilhelm-Johnen-Str., 52425 Jülich, Germany.

‡ Compound Semiconductor Technology, RWTH Aachen University, Sommerfeldstr. 18, 52074 Aachen, Germany.

¶ AIXTRON SE, Dornkaulstr. 2, 52134 Herzogenrath, Germany.

∥ Institute of Semiconductor Engineering, University of Stuttgart, Pfaffenwaldring 47, 70569 Stuttgart, Germany.

ABSTRACT: Developing electronic devices capable of emulating biological functions is essential for advancing brain-inspired computation paradigms such as neuromorphic computing. In recent years, two-dimensional materials have emerged as promising candidates for neuromorphic electronic devices. This work addresses the coexistence of volatile and nonvolatile resistive switching in lateral memristors based on molybdenum disulfide with silver as the active electrode. The fabricated devices exhibited switching voltages of ~0.16 V and ~0.52 V for volatile and nonvolatile operation, respectively, under direct-current measurements. They also displayed the essential synaptic functions of paired-pulse facilitation and short- and long-term plasticity under pulse stimulation. The operation mechanism was investigated by *in-situ* transmission electron microscopy, which showed lateral migration of silver ions along the molybdenum disulfide between electrodes. Based on the experimental data, a macroscopic semi-classical electron transport model was used to reproduce the current-voltage characteristics and support the proposed underlying switching mechanisms.

Emulating the brain and its biological functions with circuits based on conventional complementary metal-oxide semiconductor (CMOS) technology requires complex circuitry, which leads to high energy consumption.[1,2] In this scenario, resistive switching (RS) devices are promising candidates for implementing the desired artificial neurons and multifunctional synaptic devices in neuromorphic hardware. Among the wide variety of RS devices, electrochemical metallization (ECM) memristors based on two-dimensional (2D) materials exhibit particularly promising features as vertical and lateral memristors.[3–5] ECM memristors are characterized by the formation and rupture of metallic conductive filaments (CFs) under an applied electrical field.[6–8] These CFs usually originate from the migration of active metal ions such as silver (Ag) or copper (Cu).[9] Up to date, CFs composed of Ag ions have been visualized in several device architectures.[10–14] ECM memristive devices based on layered 2D materials (2DMs), including molybdenum disulfide ($MoS_2$), have demonstrated low energy consumption of approximately 10 fJ, low switching voltages of < 0.3 V, and RS in sub-nanometer thicknesses.[3,15–20] However, the growth and evolution of metallic CFs in 2DM ECM memristors rely on random defects,[10] which often lead to high cycle-to-cycle and device-to-device variability.

Lateral 2DM-based devices, in which 2DMs are placed on a dielectric substrate and contacted by metal electrodes, have been extensively studied as RS devices. [3,11,21–28] However, their switching mechanisms for volatile and nonvolatile operation (and their coexistence) remain largely unexplored. In fact, the coexistence of both RS modes by tuning the measurement system's current compliance (CC) has been only shown for vertical devices based on metal oxides and hexagonal boron nitride. [29–34]

Here, we investigate the coexistence of volatile and nonvolatile RS in lateral $MoS_2$-based memristors using Ag as the filament-forming electrode. We show the transition between volatile and nonvolatile switching modes by tuning the CC of direct-current (DC) electrical measurements. Furthermore, we demonstrate the coexistence of both modes under pulsed voltage stress (PVS). We also show short-term (STP) and long-term plasticity (LTP) in response to different pulsed stimulation and analyze the RS mechanisms for both volatile and nonvolatile switching modes. The growth and stability of the CFs depend on several factors, such as the filament morphology and the applied electrical field.[9,10,35] Therefore, we conducted *in-situ* transmission electron microscopy (TEM) images to analyze Ag migration along the $MoS_2$. Finally, we applied a semi-classical electron transport model that reproduces the experimental current-voltage (*I-V*) characteristics and supports the formation of metallic Ag CFs as the switching mechanism.

Our lateral devices comprise multilayer $MoS_2$ contacted by palladium (Pd) and active Ag electrodes with a sub-micron channel length (see schematic in Figure 1a). First, $MoS_2$ grown by metal-organic chemical vapor deposition (MOCVD) on a 2" sapphire wafer was wet transferred onto a $SiO_2$/Si substrate.[36] The Pd electrodes were defined using optical lithography, and the Ag ones were defined using a laser writer. The Ag electrodes were covered *in-situ* with a 50 nm aluminum (Al) electron-beam-evaporated (e-beam) capping layer to avoid tarnishing.[22] Reactive ion etching (RIE) was carried out to pattern the $MoS_2$ channels. More details about the device fabrication are available in the Materials and Methods section in the Supporting Information (SI), and a schematic process flow with each step is displayed in Supplementary Figure S1.

The devices were characterized by high-resolution TEM (HRTEM). The layered structure of the $MoS_2$ in a device in Figure 1b shows an interlayer distance of $6.4 \pm 0.05$ Å, in line with values reported in the literature.[37,38] The atomic force microscopy (AFM) image in Figure 1c reveals a polycrystalline nature of the as-grown MoS$_2$ on 2" sapphire. Additional AFM confirms the thickness of ~4.4 nm of the MoS$_2$ film (Supplementary Figure S2a). In addition, we conducted Raman spectroscopy measurements of the as-grown MoS$_2$ on sapphire and after transfer on SiO$_2$/Si. The extracted $E^1_{2g}$ and $A_{1g}$ peaks of MoS$_2$ coincide with the literature values for more than four layers or the bulk, which is in agreement with the AFM and HRTEM data (see Figure S2b).[39,40] Furthermore, the photoluminescence peak position of 1.82 eV corroborates the multilayered nature of the MoS$_2$ and the good crystal quality of the transferred material (Figure S2c-d).[41–43] Figure 1d shows a top-view optical microscopy image of a fabricated device. A lamella was cut from a device by focused ion beam (FIB) milling, and a cross-sectional TEM image was taken to verify the integrity of the lamella. The channel length between the metal electrodes was ~250 nm in this specific device. For the *in-situ* TEM measurements, the channels of the devices were covered with an additional 80 nm e-beam-evaporated aluminum oxide (Al$_2$O$_3$) film to protect the MoS$_2$ from the ion beam during the FIB lamella preparation. Moreover, the Al$_2$O$_3$ layer prevents direct contact between the platinum (Pt) protection layer used in FIB and both the electrodes and MoS$_2$ layers, which would otherwise lead to a short circuit.

We conducted DC electrical measurements on the Pd/MoS$_2$/Ag/Al devices with CC values varying from 100 nA to 1 mA. Volatile RS was observed for CC values from 100 nA up to 100 μA (Figure 2a). Figure 2b shows ten *I-V* curves with a 1 μA CC measured in subsequent voltage sweeps in the positive polarity and plotted in logarithmic scale (see linear scale in SI Figure S3a). The voltage bias was applied to the Ag active electrode, while the Pd electrode was grounded. The arrows indicate the voltage sweep direction. Initially, the device was in a high-resistance state (HRS) until the SET transition occurred at an on-threshold voltage of $V_{t,on}$ = 0.19 V, when it switched to its low-resistance state (LRS). During the backward sweep, the device switched back to the HRS at a hold voltage of $V_{hold}$ = 0.087 V. We plotted $V_{t,on}$ and $V_{hold}$ in histograms, and the data were fitted with Gaussian distributions that show reasonably low standard deviations ($\sigma$) of 0.04 V and 0.03 V, respectively (Figure S4). The mean $V_{t,on}$ of 0.16 V is the lowest reported among similar lateral MoS$_2$-based RS devices (see Table 1).[11,12,22,23,25] Figure 2c shows the normalized cumulative distribution functions (CDFs) of $V_{t,on}$ and $V_{hold}$ for 60 consecutive volatile cycles, with a standard error of 3.9% and 3.1%, respectively (see Figure S3b for the corresponding I-V curves).

Nonvolatile bipolar RS was observed for CC values above 500 µA (Figure 2d). We then performed ten consecutive I-V sweeps with 1mA CC, with the first sweep marked in red (Figure 2e). The extracted mean SET and RESET voltages of these switching cycles were $V_{SET}$ = 0.26 V and $V_{RESET}$ = -0.36 V. The inset in Figure 2e shows the DC endurance of a single device for ~860 consecutive nonvolatile RS cycles with an average HRS-to-LRS resistance ratio ($R_{HRS}/R_{LRS}$) of more than $10^4$. Figure 2f displays the corresponding cumulative voltage distributions with a mean $V_{SET}$ and $V_{RESET}$ of ~0.52 V and ~-0.61 V, and standard deviations of $\sigma$ = 0.3 V and $\sigma$ = 0.4 V, respectively. We further demonstrated that not only can we observe volatile and nonvolatile switching but also, we can control the transition between both in the lateral MoS$_2$ devices solely by modifying the CC value (see Figure S5).

We conducted PVS tests with different pulses to observe the volatile and nonvolatile dynamic responses of our devices. Figure 3a shows the volatile transient current response of a device to a 1 µs voltage pulse of 6 V. The switching time of $t_{on}$ = 250 ns is the time required to reach 90% of the ON current of $I_{on}$ = 12.5 µA. The relaxation time of $t_r$ = 131 ns is defined as the time needed to reach 10% of the difference between $I_{on}$ and the OFF current ($I_{off}$) when switching back to the HRS.[22] At the end of the voltage pulse, the device spontaneously relaxes to its initial OFF state. Figure 3b, in contrast, shows the transient nonvolatile output current response to a 0.02 s voltage pulse of $V_{SET} = 3.5$ V and a 0.05 s voltage pulse of $V_{RESET} = -5$ V. Read voltage pulses of $V_{read} = 0.2$ V and 0.1V were applied to measure the resistances in the LRS and HRS, respectively. After the SET pulse, the current did not return to its initial OFF-state but remained in the ON-state. The current pulse levels for the nonvolatile switching are over 100 µA, which matches the levels applied in the DC CC-dependent measurements. The $R_{LRS}$ after the SET pulse was ~1.4 kΩ, whereas the $R_{HRS}$ was ~200 MΩ.

STP and LTP were investigated by applying a series of consecutive voltage pulses. First, a train of 40 consecutive 1 V / 50 ns pulses with a 50 ns period between them was applied and the transient response of our device recorded. The current response in Figure 3c shows a gradual current increase with each pulse, which saturates at 13 nA after about 18 pulses. The current decreases slightly after each pulse and decays to its initial state after the final voltage pulse. The fast relaxation to the initial OFF-state with no memory retention and a relaxation time of $t_r = 2.2$ µs, emulates STP, i.e., a forgetting process observed in biological synapses.[44] We also achieved LTP in the same device by tuning the programming parameters so the device current increased to a higher value, triggering the non-volatile regime. Figure 3d shows the device current response to a train of 20 consecutive pulses of an increased voltage of 2 V and a much longer duration of 2 ms. A magnification of the current response in the inset shows a gradual current increase after each applied pulse, resembling the biological synaptic function known as paired-pulse facilitation (PPF). After the final pulse, the current decayed to an intermediate ON-state, in contrast to the previous experiment with 40 smaller, shorter pulses. The retention time of the LRS was recorded for over ~100 s, more than the time displayed in Figure 3d (see SI Figure S6). This device behavior aligns with the LTP concept, as the LRS lasts for more than a few tens of seconds.[44–46] Controlling the transition from STP to LTP by tuning the programming parameters of the pulse train is in line with previous works on SiO$_2$ and Ag$_2$S-based memristors.[45,47]

We further investigated the current conduction mechanisms of our memristive devices. We plotted the *I-V* curves from Figures 2b (volatile RS) and 2e (nonvolatile) in double-logarithmic scale in Figures 4a and 4b. In the HRS, the *I-V* curves exhibit a square-law dependence, indicating space-charge-limited conduction (SCLC),[48,49] which has been previously reported in two-terminal lateral polycrystalline MoS$_2$-based devices.[50] In the LRS, the linear *I-V* dependence indicates ohmic conduction. This transition from SCLC in the HRS to ohmic transport in the LRS supports the proposed formation of Ag CFs. Temperature-dependent *I-V* measurements confirmed these transport mechanisms (Figure S7). In the HRS, the current increases with increasing temperature, which is characteristic of transport through the MoS$_2$ channel and SCLC.[50] In the LRS, the current decreases with increasing temperature. Such a negative temperature coefficient of resistivity is characteristic of transport through metallic CFs.[51] Both volatile and nonvolatile RS mechanisms exhibit similar transport characteristics, suggesting the formation of a metallic CF in both cases. However, the applied CC plays a crucial role in determining the RS behavior. We propose that low CC values lead to volatile RS with a weak Ag filament that self-ruptures upon removing the applied voltage. In contrast, higher CC values lead to nonvolatile RS because more Ag ions diffuse into the channel and create a stronger Ag CF. Therefore, we deduce that Ag ions diffuse into the channel driven by the lateral electric field and eventually form continuous Ag filaments.

We conducted *in-situ* TEM on lateral Pd/MoS$_2$/Ag devices. Figure 5a shows the typical TEM results from a fresh device in the HRS (without any electrical measurement before), including a bright-field TEM image, high-angle annular dark-field (HAADF) images and the corresponding energy-dispersive X-ray spectroscopy (EDXS) elemental maps of Mo, S, and Ag. No traces of Ag are observed between the electrodes within the MoS$_2$ channel. This observation is confirmed in Figure S8a, where the measured EDXS line profile for Ag is flat across the multilayered MoS$_2$. The Pd electrode of the device lamella was then electrically biased *in-situ* in the TEM while the Ag electrode was grounded (see Figure S8c). The cross-sectional TEM image in Figure 5b shows the lamella device after *in-situ* nonvolatile switching to the LRS. The arrows on the TEM image indicate three different regions in which HAADF images and their corresponding EDXS maps were taken (Figure 5c-e). The presence of Ag is clearly visible in these elemental maps. Further, the EDXS line profile of Ag coincides with those of Mo and S, proving the presence of Ag within the MoS$_2$ channel (Figure S8b). This confirms the formation of an Ag CF following the SET transition. These findings experimentally prove Ag ion migration driven by an applied lateral electrical field, in line with previous reports in the literature.[12,14,23]

We investigated lateral memristive devices with polycrystalline 2D multilayer MoS$_2$ channels with sub-micrometer lengths. We demonstrated the coexistence of volatile and nonvolatile RS in the same devices, attainable with both DC and PVS. Our devices display repeatable RS with low switching voltages for both volatile (0.16 V) and nonvolatile (0.52 V) operation. The volatile behavior may be utilized in artificial neurons. We also experimentally demonstrated PPF, STP and LTP behavior in our devices, showcasing their potential as artificial synapses for neuromorphic systems. We investigated the current conduction mechanisms in the HRS and LRS, concluding that SCLC and ohmic transport are the dominant mechanisms, respectively. This behavior can be explained by the formation of Ag conductive filaments, which was confirmed via *in-situ* TEM switching experiments. The flexibility of volatile and nonvolatile switching in our lateral memristors can be used to implement both synaptic devices and artificial neurons for future neuromorphic systems.

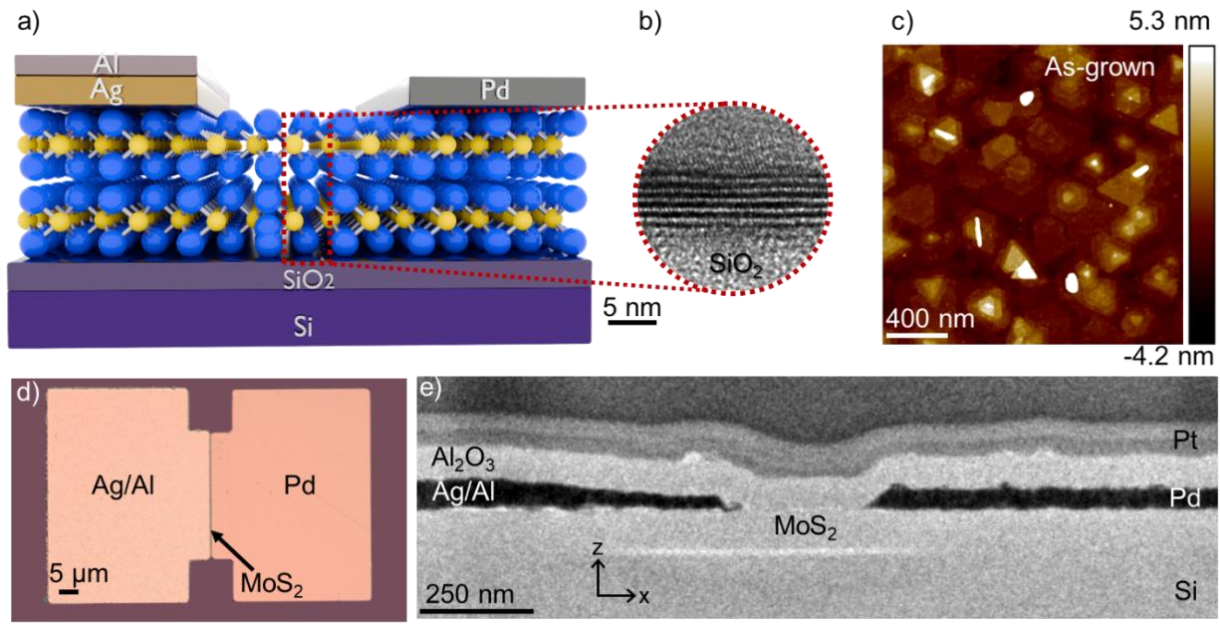

**Figure 1.** Device structure and material characterization. (a) Cross-section schematic of the lateral MoS$_2$-based memristor. (b) HRTEM image of the multilayer MoS$_2$ film between the metal electrodes. (c) AFM image of the as-grown MoS$_2$ on sapphire. (d) Top-view optical microscopy image of a fabricated lateral device. (e) TEM image showing the cross section of a device. The channel length was approximately 250 nm. Al$_2$O$_3$ was e-beam-evaporated to protect the MoS$_2$ during FIB.

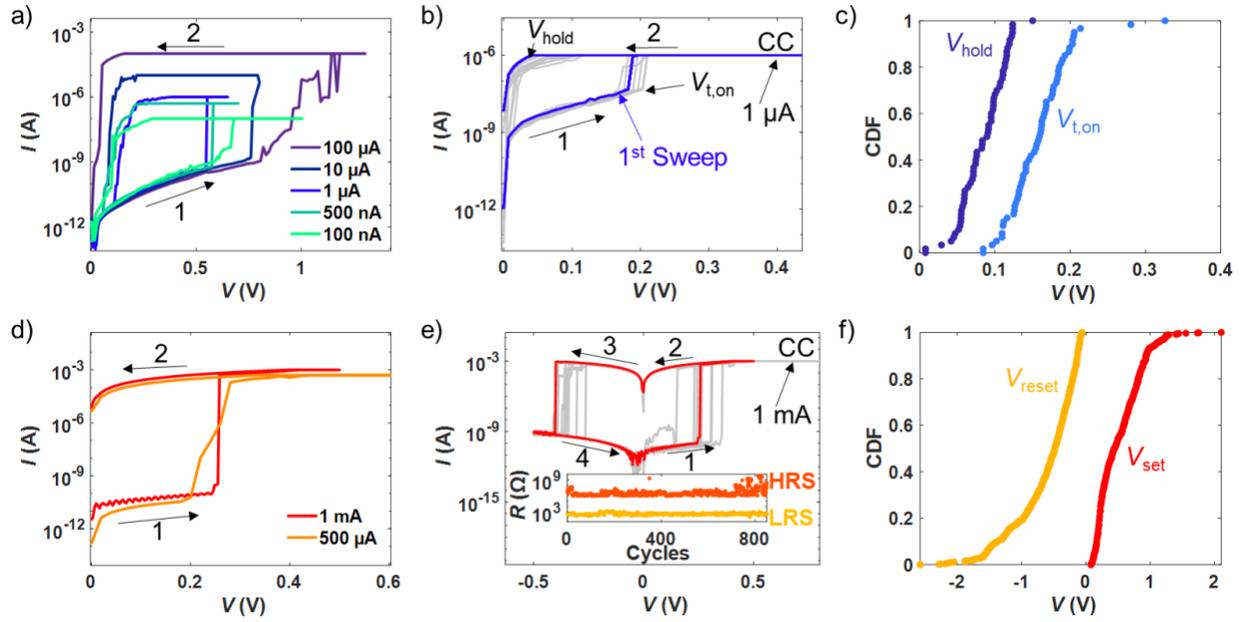

**Figure 2.** DC electrical performance of the lateral MoS$_2$-based devices. (a) Volatile RS at different CC values from 100 nA up to 100 µA. (b) 10 subsequent volatile switching curves recorded with positive voltage polarity for 1 µA CC. The arrows indicate the voltage sweeping directions, with the first sweep marked in blue. (c) CDFs of the $V_{t,on}$ and the $V_{hold}$ of 60 consecutive volatile I–V curves. (d) Nonvolatile RS at 500 µA and 1 mA CC. (e) 10 consecutive nonvolatile I-V curves for 1 mA CC. The inset shows the resistance values for ~860 nonvolatile consecutive cycles. (f) CDFs of the $V_{SET}$ and $V_{RESET}$ of ~860 nonvolatile cycles.

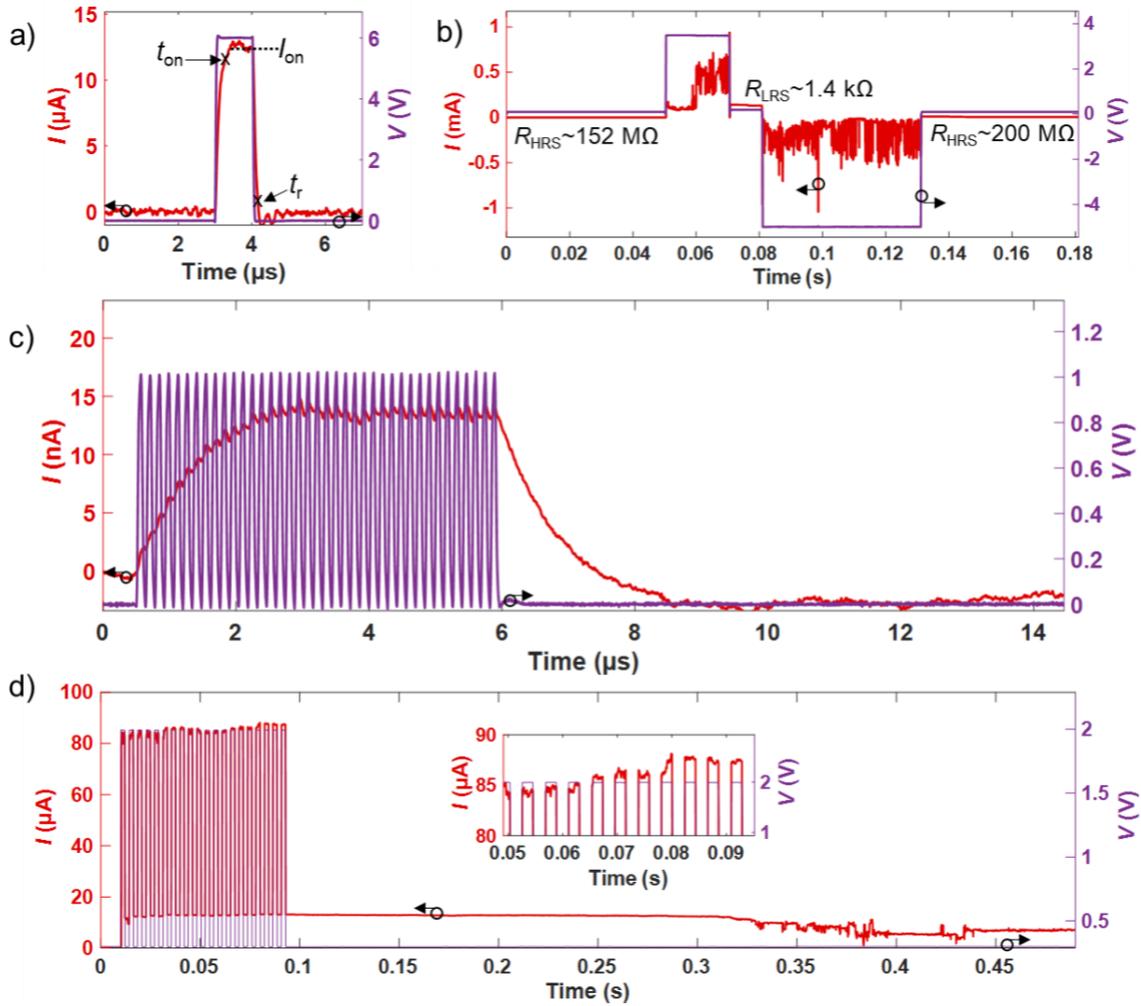

**Figure 3.** Current response over time of a lateral memristive device upon PVS. (a) Voltage pulse of 6 V for 1 µs showing volatile RS. (b) Nonvolatile RS upon PVS with a 3.5 V / 20 ms SET pulse and a - 5 V / 50 ms RESET pulse. (c) Dynamic response of a device during a sequence of 40 voltage pulses of 1 V for 50 ns showing short-term potentiation with forgetting upon stimuli removal. (d) Train of 20 consecutive 2 V / 2 ms pulses showing long-term plasticity. The read voltage applied between each pulse was 0.3 V. The inset shows the gradual increase in the current after each applied pulse.

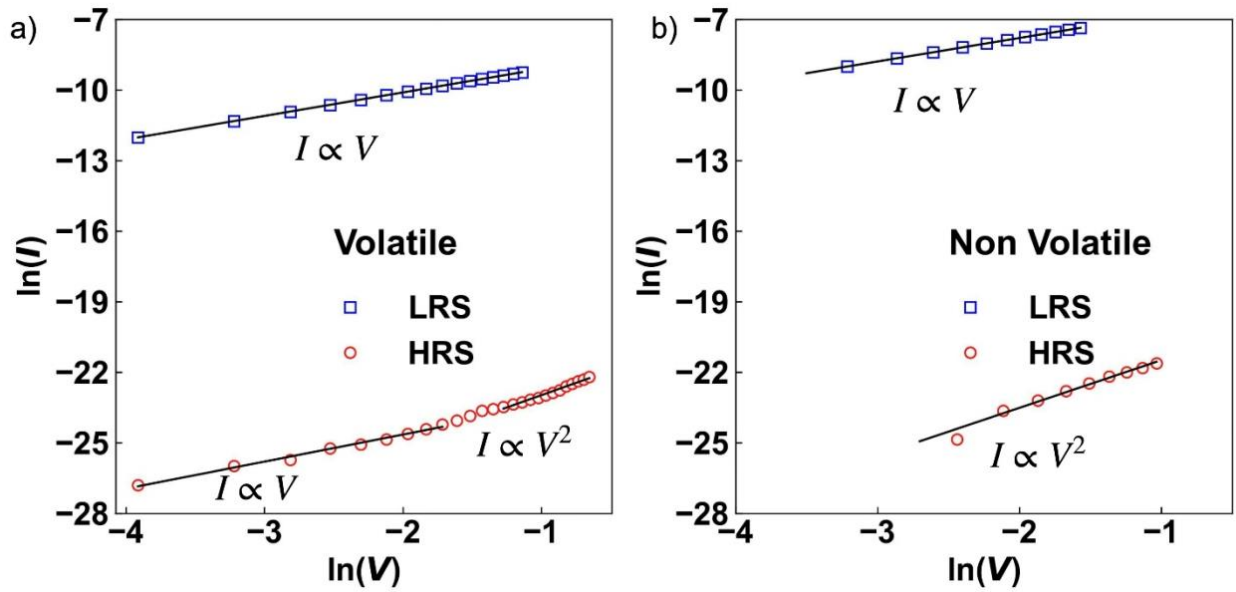

**Figure 4.** *I-V* characteristics of (a) a volatile and (b) a nonvolatile operation. In both cases, HRS shows SCLC, while LRS shows ohmic transport.

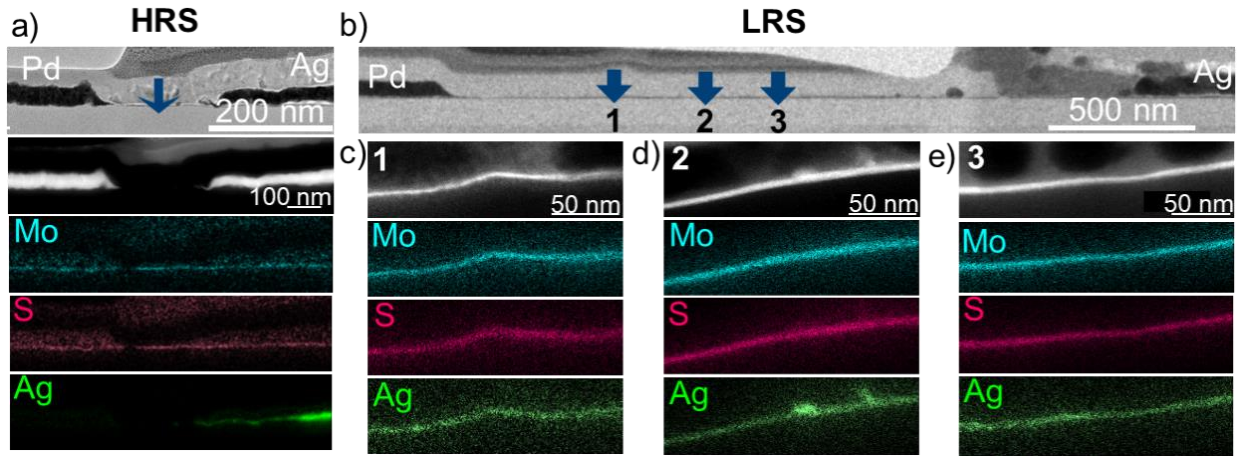

**Figure 5.** *In-situ* TEM measurements for the resistive switching mechanism analysis. (a) Cross-sectional bright-field TEM image, HAADF image and the corresponding EDXS elemental maps of Mo, S, and Ag from a lamella in the HRS before any electrical measurement. The blue arrow denotes the region where the EDXS elemental maps were acquired. (b) Cross-section TEM image of a lamella after nonvolatile RS in the LRS. (c-e) EDXS elemental maps of the lamella in the LRS taken on the regions denoted as 1, 2, and 3 in (b), respectively.

**Table 1.** Comparison of our work with other similar 2D-based literature.

| Ref. | Device structure | Material | Thickness | Forming voltage (V) | Volatile ($V_{t,on}$) | Nonvolatile ($V_{SET}$) | In situ of switching mechanism |
|---|---|---|---|---|---|---|---|
| **This work** | **Lateral** | **MOCVD-grown MoS$_2$** | **4.4 nm** | **Forming-free** | **O (0.16 V)** | **O (0.52 V)** | **O** |
| 25 | Lateral | CVD-grown MoS$_2$ | Monolayer | ~20 V | - | O (3.5-8.3 V) | O |
| 3 | Lateral | CVD-grown MoS$_2$ | Monolayer | ~4 V | O (1.2 V) | - | O |
| 23 | Lateral | Mechanically exfoliated MoS$_2$ | <6 layers | 1.8 V | O (0.35 V) | - | - |
| 12 | Lateral | Mechanically exfoliated MoS$_2$ | 4-30 layers | ~3.8 V | - | O (2 V) | O |
| 11 | Lateral | CVD-grown MoS$_2$ | Monolayer | Forming-free | - | O (10-20 V) | O |
| 10 | Lateral | Exfoliated MoS$_2$ nanoflakes | >1 layer | - | - | O (~6 V) | O |
| 52 | Vertical | CVD-grown vertically aligned MoS$_2$ | ~23 nm | Forming-free | O (~0.35 V) | - | - |

ASSOCIATED CONTENT

The Supporting Information is available free of charge.

Materials and methods including MoS$_2$ growth process, lateral device fabrication, material and device characterization, and electrical measurements; MoS$_2$ material characterization; Schematic of the fabrication process; Volatile *I-V* curves in linear and log scale; Histogram and Gaussian fit for the on-threshold and hold voltages; Transition from volatile to nonvolatile back and forth; LRS retention time; Lamella and setup preparation for *in-situ* TEM; Benchmarking with previous works (PDF)

AUTHOR INFORMATION

**Corresponding Author**

Max C. Lemme − Chair of Electronic Devices, RWTH Aachen University, Otto-Blumenthal-Str. 25, 52074 Aachen, Germany; AMO GmbH, Advanced Microelectronic Center Aachen, Otto-Blumenthal-Str. 25, 52074 Aachen, Germany. *Email: max.lemme@eld.rwth-aachen.de

**Author Contributions**

M.C.L and A.D. conceived and designed the project. S.C. carried out the material characterization, mask design, device fabrication, and electrical measurements. B.C. supported S.C. with the AFM measurements. H.K, A.V. and M.H. provided the polycrystalline MoS$_2$. K.R., J.J. performed *in-situ* TEM and EDXS experiments. M.D.G. supported the analysis of the electrical data and the writing. S.C. wrote the initial manuscript, and all authors cowrote it. All authors have given approval to the final version of the manuscript.

**Notes**

The authors declare no competing financial interest.

ACKNOWLEDGMENTS

Financial support from the German Federal Ministry of Education and Research (BMBF) within the projects NEUROTEC 2 (No. 16ME0399, 16ME0400, and 16ME0403) and the Clusters4Future NeuroSys (No. 03ZU1106AA and 03ZU1106AD) is gratefully acknowledged. We acknowledge funding from the European Union's Horizon Europe research and innovation program (via CHIPS-JU) under the project ENERGIZE (101194458). A.D. acknowledges support by the German Research Foundation through the Emmy Noether Programme (506140715). We also acknowledge support by the Spanish Government through the research projects CNS2023-143727 RECAMBIO, PID2023-150162OB-I00 ADAGE, and TED2021-129769B-I00 FlexPowHar funded by MCIN/AEI/10.13039/501100011033 and the European Union NextGenerationEU/PRTR.

# Supporting Information

# Volatile and Nonvolatile Resistive Switching in Lateral 2D Molybdenum Disulfide-Based Memristive Devices

*Sofía Cruces[#], Mohit D. Ganeriwala[†], Jimin Lee[#], Ke Ran[¥,§,Ω], Janghyun Jo[Ω], Lukas Völkel[#], Dennis Braun[#], Bárbara Canto[¥], Enrique G. Marín[†], Holger Kalisch[‡], Michael Heuken[‡,¶], Andrei Vescan[‡], Rafal Dunin-Borkowski[Ω], Joachim Mayer[§,Ω], Andrés Godoy[†], Alwin Daus[∥], and Max C. Lemme[#,¥*]*

[#] Chair of Electronic Devices, RWTH Aachen University, Otto-Blumenthal-Str. 25, 52074 Aachen, Germany.

[†] Department of Electronics and Computer Technology, Facultad de Ciencias, Universidad de Granada, 18071, Granada, Spain.

[¥] AMO GmbH, Advanced Microelectronic Center Aachen, Otto-Blumenthal-Str. 25, 52074 Aachen, Germany.

[§] Central Facility for Electron Microscopy, RWTH Aachen University, Ahornstr. 55, 52074, Aachen, Germany.

[Ω] Ernst Ruska-Centre for Microscopy and Spectroscopy with Electrons (ER-C), Forschungszentrum Jülich GmbH, Wilhelm-Johnen-Str., 52425 Jülich, Germany.

[‡] Compound Semiconductor Technology, RWTH Aachen University, Sommerfeldstr. 18, 52074 Aachen, Germany.

[¶] AIXTRON SE, Dornkaulstr. 2, 52134 Herzogenrath, Germany.

[∥] Institute of Semiconductor Engineering, University of Stuttgart, Pfaffenwaldring 47, 70569 Stuttgart, Germany.

* Corresponding author. Email: max.lemme@eld.rwth-aachen.de (MCL)

**This file includes:**

Materials and Methods

Supporting Figures S1-S8

Supporting References

**Materials and Methods**

**Metal–organic chemical vapor deposition (MOCVD) of MoS$_2$**

Highly uniform multi-layered molybdenum disulfide (MoS$_2$) has been epitaxially grown in a commercial AIXTRON planetary hot-wall reactor in 10 × 2" configuration on c-plane sapphire substrates (with a 0.2° offcut toward m-plane). First, a substrate desorption at 970 °C in a pure H$_2$ atmosphere has been performed. Afterward, growth has been carried out at a substrate temperature of 750 °C for 24 h, with nitrogen as the carrier gas and at a pressure of 30 hPa. The precursor flows have been set to 19 nmol/min for molybdenum hexacarbonyl (MCO) and 19 μmol/min for di-tert-butyl sulfide (DTBS).

**Device Fabrication**

MoS$_2$ grown on 2" sapphire by MOCVD was transferred onto 2 × 2 cm$^2$ Si chips covered with 275 nm thermal SiO$_2$ thermally grown on Si wafers with p-doping of 1–10 Ω·cm. Poly(methyl methacrylate) (PMMA) was spin-coated on top of the MoS$_2$ before being released from the sapphire substrate in deionized water.[1] For both electrodes the AZ5214E JP photoresist from Merck Performance Materials GmbH was used. The palladium (Pd) electrodes were defined using optical contact lithography with an EVG 420 Mask Aligner and the silver (Ag) ones with a Microtech LW405C laser writer. The asymmetric Pd (50 nm) and Ag (50 nm)/aluminum (Al) (50 nm) electrodes were deposited via electron-beam evaporation (e-beam) in a tool from FHR Anlagenbau GmbH and subsequently lifted in acetone at room temperature. Finally, the channels were patterned via CF$_4$ (20 sccm)/O$_2$ (10 sccm) reactive ion etching (RIE) in an Oxford Instruments Plasma Lab System 100 tool. The resist was stripped in acetone at 60 °C for 1 h. For the in-situ experiments, an extra 80 nm aluminum oxide (Al$_2$O$_3$) layer was deposited via e-beam in a tool from FHR Anlagenbau GmbH. The Al$_2$O$_3$ layer is required not only to protect the MoS$_2$ from the ion beam during the focused ion beam (FIB) preparation but also to avoid direct contact of the platinum (Pt) protection layer with both electrodes and $MoS_2$ layers. If the Pt protection layer is directly in contact with both electrodes, current would bypass the $MoS_2$ layers, and a short circuit would be observed. For the same reason, the Pt protection layer was removed in the center of the lamella (on top of the $MoS_2$ and the $Al_2O_3$ layer as shown in Figure S8c).

**Material and Device Characterization**

Optical microscope images were recorded with a Leica INM100 microscope and a Keyence 3D Laser Scanning microscope VK- X3000. Raman and photoluminescence (PL) measurements were performed with a WiTec alpha300R Raman spectrometer with an excitation laser wavelength of 532 nm and 1 mW laser power. TEM specimens were prepared by FIB milling using an FEI Strata400 system with gallium (Ga) ion beam. Transmission electron microscopy (TEM) analysis was conducted with a JEOL JEM F200 instrument at 200 kV. *In-situ* TEM was conducted with a FEI Titan G2 80-200 ChemiSTEM microscope at 200 kV equipped with an XFEG, a probe Cs corrector and a super-X energy-dispersive X-ray spectroscopy (EDXS) system. A voltage bias was applied *in-situ* to the TEM lamella in the TEM using a Nanofactory STM-TEM specimen holder connected to an external "Keithley 2602A" source meter. Atomic force microscopy (AFM) measurements were conducted with a Dimension Icon AFM from Bruker Instruments using a tip with 26 N/m as force constant and the frequency variation between 200 and 400 kHz (OTESPA) in tapping mode in air and at room temperature.

**Electrical Measurements**

Electrical measurements were performed in a LakeShore probe station connected to a semiconductor parameter analyzer (SPA) "Keithley 4200A-SCS" with two source measure unit (SMU) cards "Keithley 4200-SMU", each connected to a preamplifier "Keithley 4200-PA" from Tektronix. A voltage bias was applied to the Ag/Al active electrode, and the Pd electrode was grounded. Current-voltage *(I–V)* measurements were conducted by sweeping the voltage from 0 V to a positive maximum voltage $V_{max}$ and back to 0 V. For the nonvolatile measurements the same was applied in the negative polarity for resetting the devices. The current is limited by an external current limiter within the semiconductor parameter analyzer. Pulse experiments were performed by supplying a voltage to the Ag/Al active electrode (channel 1) and measuring the output current over time in channel 2 (Pd electrode). Temperature-dependent measurements were conducted for each temperature in two steps. First, 10 measurement points at a constant read voltage of 100 mV were collected to extract the device resistance. Second, a forward sweep from −0.2 V to 0.2 V was applied.

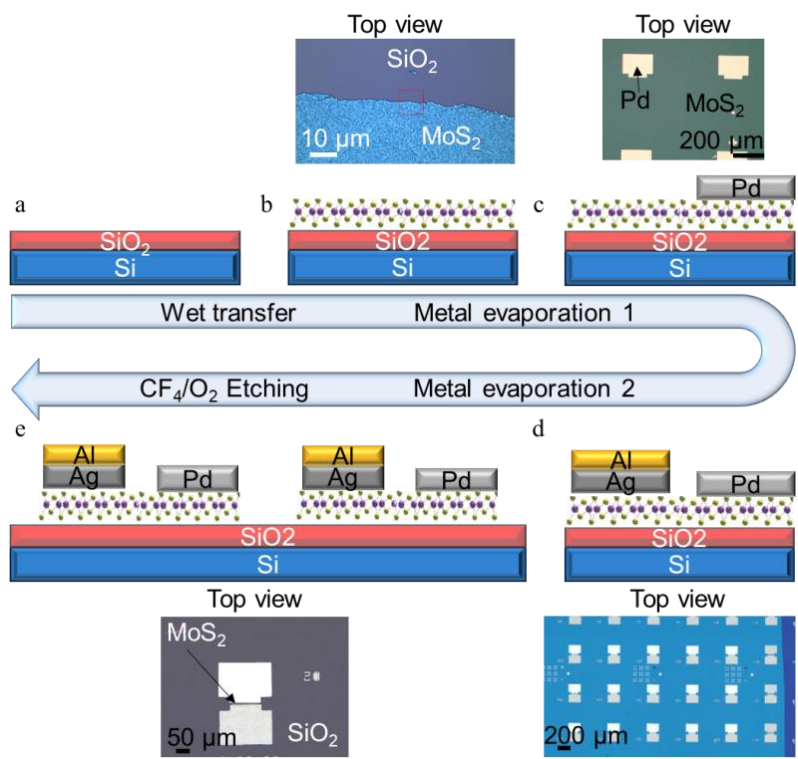

**Supplementary Figure S1. Fabrication process flowchart.** (a) Clean $SiO_2$/Si substrate. (b) Wet transfer of $MoS_2$ film from sapphire substrate to $SiO_2$/Si. Top view optical microscopy (OM) image of $MoS_2$ on $SiO_2$ after transfer. (c) Electron beam (e-beam) evaporation of the Pd metal contact after photolithography with negative photoresist and lift-off. The OM image shows a top view of the Pd electrode on $MoS_2$. (d) E-beam evaporation of Ag and Al on top as capping layer after photolithography with negative photoresist and lift-off. The OM shows a top view of multiple devices with both electrodes on top of the $MoS_2$ film. (e) Reactive ion etching of the $MoS_2$ active areas with $CF_4/O_2$ after photolithography with positive photoresist followed by resist stripping in acetone. The OM images show a device after photoresist stripping and a close up to the gap between electrodes.

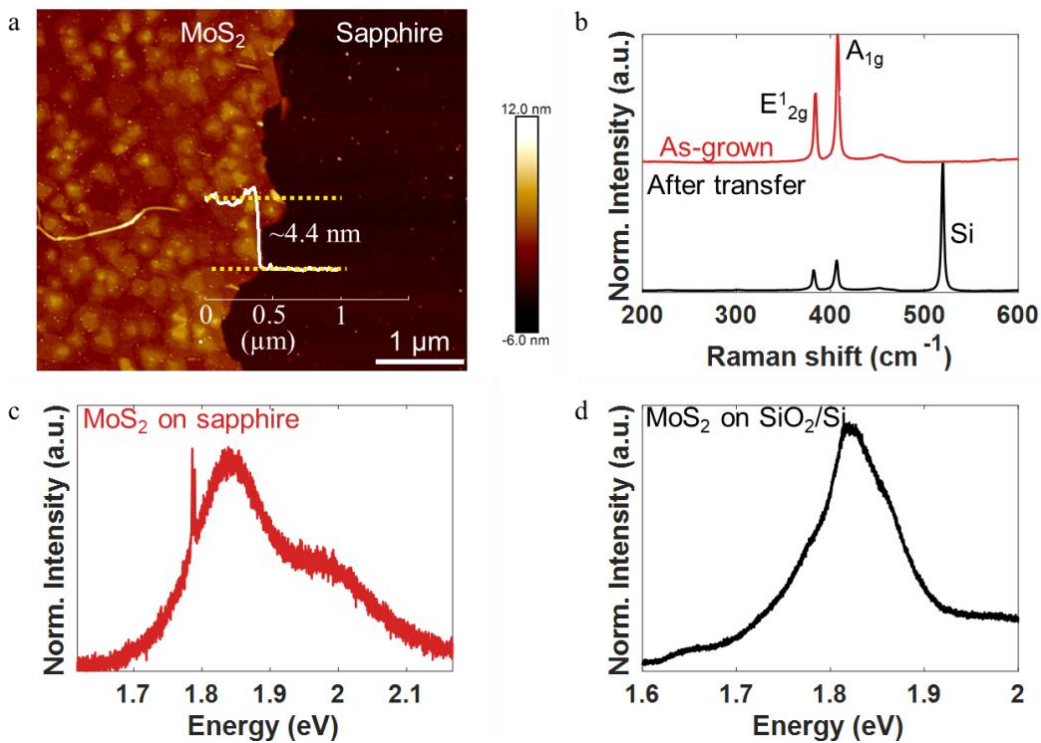

**Supplementary Figure S2. MoS₂ material characterization.** (a) Atomic force microscopy (AFM) image of the as-grown $MoS_2$ on sapphire. The inset height profile shows a thickness of ~4.4 nm. (b) Raman spectra of as-grown $MoS_2$ on sapphire and after transfer on a $SiO_2$/Si substrate. The extracted peaks for the as-grown $MoS_2$ on sapphire were 383.9 cm$^{-1}$ and 407.9 cm$^{-1}$ for the $E^1_{2g}$ and $A_{1g}$ peaks, respectively. In the case of the transferred $MoS_2$ onto an $SiO_2$/Si substrate, the obtained values were 382.4 cm$^{-1}$ and 406.5 cm$^{-1}$ for the $E^1_{2g}$ and $A_{1g}$ peaks, respectively. All the extracted values match those previously reported for four layers or bulk.[2,3] PL spectrum of (c) as-grown $MoS_2$ on sapphire showing both PL peaks and (d) after transfer on a $SiO_2$/Si substrate. The PL peak position located at 1.82 eV is in good agreement with the theoretical value for multilayer $MoS_2$ [4–6], indicating high crystal quality of the transferred material.

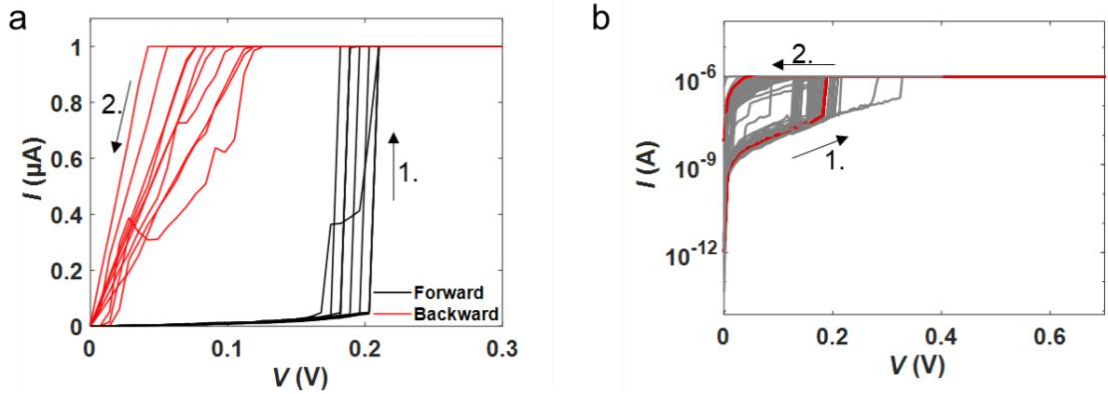

**Supplementary Figure S3. Volatile *I-V* curves on a linear and logarithmic scale.** (a) 10 consecutive volatile resistive switching (RS) curves at 1 µA CC in linear scale. The forward sweep is marked in black, and the backward sweep is marked in red. (b) 60 consecutive volatile RS cycles at 1 µA CC in logarithmic scale. The first sweep is marked in red.

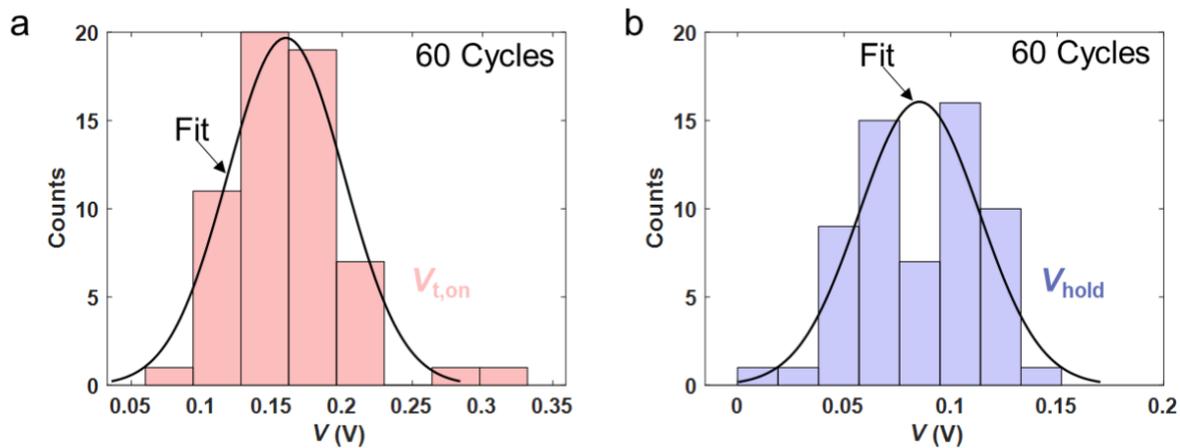

**Supplementary Figure S4. Statistical analysis of switching voltages for $V_{t,on}$ and $V_{hold}$.** (a) Histogram and Gaussian fit for $V_{t,on}$ derived from 60 volatile RS data points. (b) Histogram and Gaussian fit for $V_{hold}$ derived from 60 volatile RS data points. The statistical distributions of the $V_{t,on}$ and $V_{hold}$ show low mean voltage values and low standard deviations ($\sigma$) of $0.16 \pm 0.04$ V and $0.085 \pm 0.03$ V, respectively.

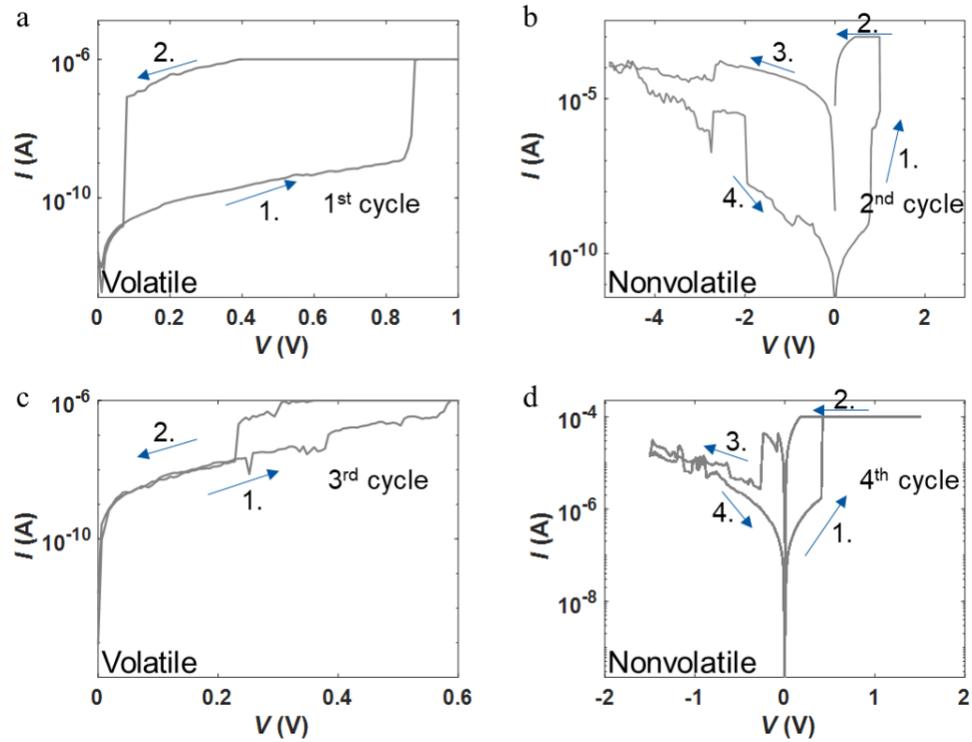

**Supplementary Figure S5. Transition from volatile to nonvolatile back and forth.** (a) One volatile threshold RS *I-V* sweep with a CC at 1 µA (1st cycle). (b) One nonvolatile RS *I-V* sweep with a CC at 1 mA after the first one in (a) showing both set and reset processes (2nd cycle). (c) One volatile threshold RS *I-V* sweep with a CC at 1 µA conducted after the nonvolatile RS cycle in (b) (3rd cycle). (d) One nonvolatile RS *I-V* sweep (4th cycle) with a CC at 1 mA. The arrows show the voltage sweep direction.

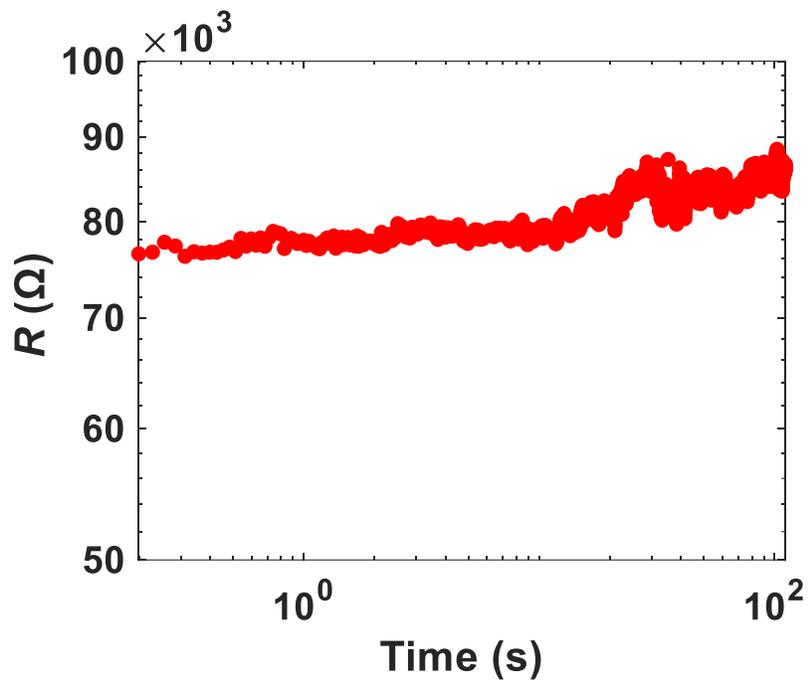

**Supplementary Figure S6: Low-resistance state (LRS) retention over 100 s after continuous pulsed stimulation.** Retention of the LRS with a DC measurement for 4096 data points.

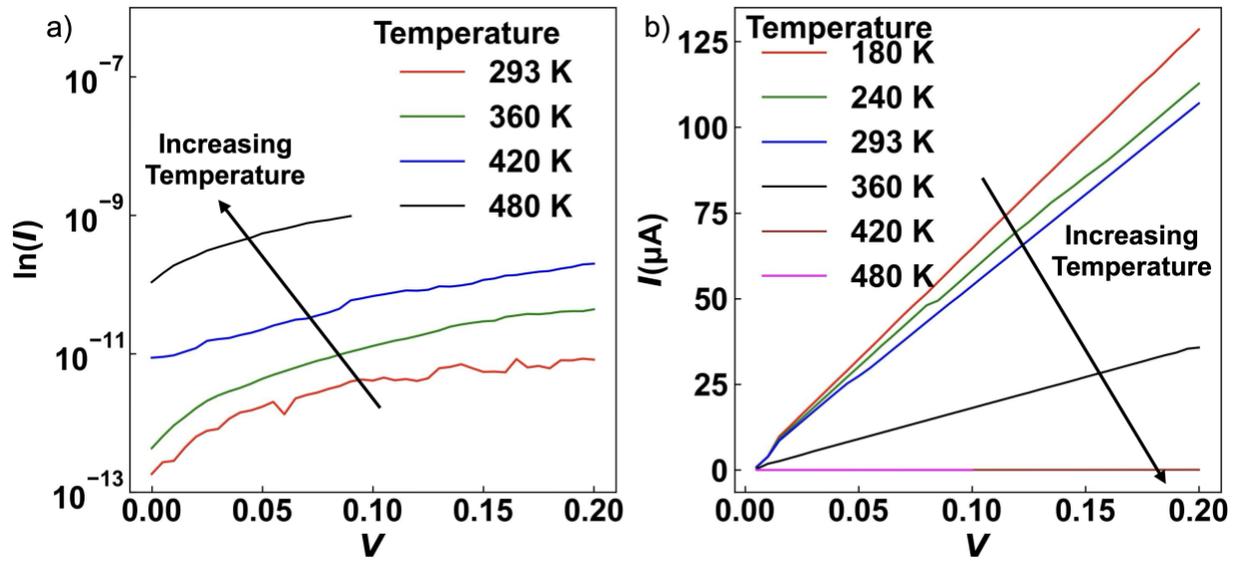

**Supplementary Figure S7. Temperature dependent measurements for the HRS and the LRS.** (a) In the HRS the current increases with increasing temperature, which is characteristic of transport through the MoS$_2$ channel and SCLC.[7] (b) In the LRS the current decreases with increasing temperature, indicating a negative temperature coefficient of resistivity which is characteristic of transport through a metallic CF.[8]

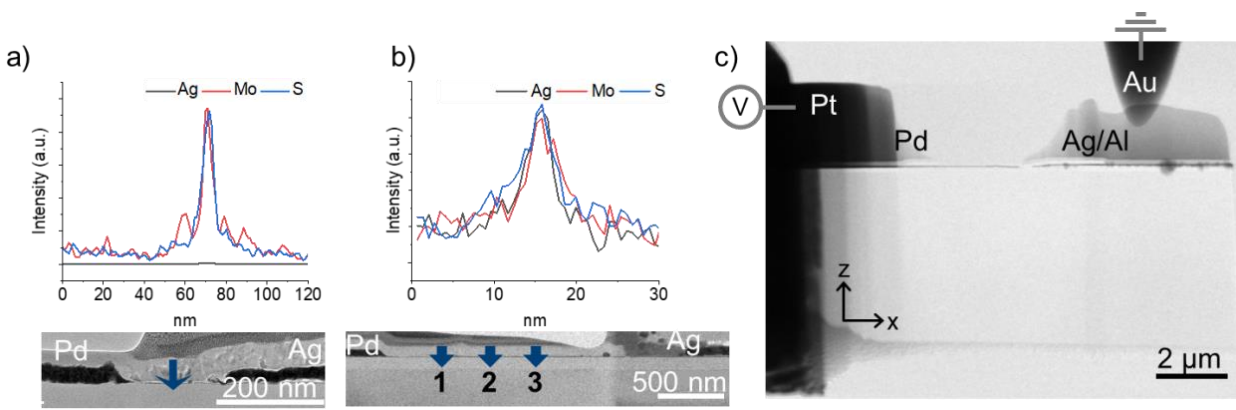

**Supplementary Figure S8. EDXS line profiles, HRTEM and lamella for *in-situ* TEM electrical measurements.** The EDXS line profiles showing the intensities of corresponding characteristic X-rays were recorded along the vertical direction pointed by the blue arrows marked in the cross-section TEM images. (a) EDXS line profile of a lamella in the HRS as shown in the TEM image, before any electrical measurement is carried out. The profile for Ag is flat, proving the absence of Ag within the $MoS_2$ and in the channel between electrodes. (b) EDXS line profile of a lamella in the LRS taken at position 1 as depicted in the TEM image. The intensity profile of Ag coincides with those from Mo and S, proving the presence of Ag within the $MoS_2$ channel. (c) TEM image of one lamella used for the *in-situ* TEM measurements. The Pd electrode was biased while the Ag active electrode was grounded.